\documentclass{aa}

\usepackage{psfig}

\newcommand{\msolyr}{$M_{\odot}$\,yr$^{-1}$}

\newcommand{\hi}{H\,{\small{\sc I}}}
\newcommand{\hii}{H\,{\small{\sc II}}}

\begin{document}


\title{The quest for hot gas in the halo of NGC\,1511}

\author{M. Dahlem\inst{1}\thanks{Present address: Australia Telescope
  National Facility, Paul Wild Observatory, Locked Bag 194, Narrabri 
  NSW 2390, Australia; email: mdahlem@atnf.csiro.au}
\and
M. Ehle\inst{2,3}
\and
F. Jansen\inst{3}
\and 
T. M. Heckman\inst{4}
\and
K. A. Weaver\inst{5}
\and
D. K. Strickland\inst{4}
}

\institute{European Southern Observatory, Alonso de Cordova 3107, 
  Vitacura, Casilla 19001, Santiago 19, Chile
\and
XMM-Newton Science Operations Centre, Apartado 50727, 28080 
  Madrid, Spain
\and
Science Operations \& Data Systems Division, Research and Scientific 
  Support Department of ESA, ESTEC, 2200 AG Noordwijk, The
  Netherlands
\and
Dept. of Physics and Astronomy, Johns Hopkins University,
3400 N. Charles St., Baltimore, MD 21218, USA
\and
NASA/Goddard Space Flight Center, Code 662, Greenbelt, MD 20771, USA
}

\offprints{Michael.Dahlem@csiro.au}

\date{Received 02 January 2003 / Accepted 05 March 2003}

\abstract{
{\it XMM-Newton} observations of the starburst galaxy NGC\,1511
reveal the presence of a previously unknown extended hot gaseous
phase of its ISM, which partly extends out of the disk plane.
The emission distribution is asymmetric, being brightest in
the eastern half of the galaxy, where also radio continuum
observations suggest the highest level of star formation.
Spectral analysis of the integral 0.2--12 keV X-ray emission 
from NGC\,1511 indicates a complex emission composition. A
model comprising a power law plus thermal plasma component,
both absorbed by foreground gas, cannot explain all details
of the observed spectrum, requiring a third spectral component
to be added. This component can be a second thermal plasma,
but other spectral models can be fitted as well.
Its X-ray properties characterize NGC\,1511 as a starburst
galaxy. The X-ray-to-infrared luminosity ratio is consistent
with this result.
Together with the X-ray data, {\it XMM-Newton} obtained 
UV images of NGC\,1511, tracing
massive stars heating the ambient gas, which is then seen in 
H$\alpha$ emission. UV, H$\alpha$ and near-infrared imagery 
suggest that NGC\,1511 is disturbed, most likely by its two 
small companions, NGC\,1511a and NGC\,1511b.
}

\maketitle

\section{Introduction}

The most recent generation of X-ray satellites, {\it XMM-Newton}
and {\it Chandra}, has a much higher sensitivity than previous X-ray
missions. This enables us to conduct observations of fainter targets 
than before and to obtain a much more detailed picture of those
known already. Increased sensitivity leads to significant progress 
in investigations of low surface-brightness emitters, such as for 
example hot gaseous halos around actively star-forming spiral 
galaxies. As part of the {\it XMM-Newton} Guaranteed Time program
we have observed the edge-on galaxy NGC\,1511. Here we report on 
the results from these observations.

The {\it XMM-Newton} observations of nearby starburst galaxies
form part of a multi-wavelength study of all phases of extraplanar 
gas in external galaxies, which is conducted in order to assess the 
importance of halos as repositories of a metal-enriched medium and 
their significance in terms of galactic chemical evolution and 
possible metal enrichment of the intergalactic medium (IGM; e.g. 
Heckman et al. 1990; Dahlem et al. 1995, 1998; Weaver et al. 2000; 
Heckman 2001). Soft X-ray surveys of nearby spiral galaxies,
including starburst systems, based on {\it ROSAT} data were 
published e.g. by Read et al. (1997) and Read \& Ponman (2001).

\begin{table}[h!]
\begin{flushleft}
\leavevmode
\caption{Salient properties of NGC\,1511}
\label{tab:n1511}
\begin{tabular}{llc}
\noalign{\hrule\smallskip}
Distance & $D$ [Mpc] & 17.5 \\
IR luminosity & log($L_{\rm IR}$) [erg s$^{-1}$] & 43.66 \\
FIR flux ratio & $S_{60}/S_{100}$ & 0.54 \\
H$\alpha$ luminosity & log($L_{\rm H\alpha}$) [erg s$^{-1}$] & 40.56 \\
Absolute B magnitude & $M_{\rm B}$ [mag] & --19.6 \\
Rotation velocity & $v_{\rm rot}$ [km s$^{-1}$] & 112 \\
Star formation rate & {\it SFR}$_{\rm IR}$ [\msolyr] & 2.7 \\
\noalign{\smallskip\hrule}
\end{tabular}
\end{flushleft}
\end{table}

NGC\,1511 is an SAa pec: \hii\ type galaxy with warm dust, 
as traced by its high far-infrared flux ratio. Its salient
properties are collected in Table~\ref{tab:n1511}\ (from
Lehnert \& Heckman 1995; scaled to $D = 17.5$ Mpc). These
values indicate that NGC\,1511 is an intermediate-mass
galaxy (more massive than M\,82, but less massive than
NGC\,253) with a high star formation rate ({\it SFR}).
Optical spectroscopy was published by Lehnert \& Heckman (1995)
and, recently, by Kewley et al. (2001), establishing its starburst
nature.
Two extraplanar gas phases were previously detected in NGC\,1511:
H$\alpha$ emission from warm ionized gas (Lehnert \& Heckman 
1995) and nonthermal radio continuum emission from cosmic-rays 
and magnetic fields (Dahlem et al. 2001). These detections, 
together with the high 
surface brightness of tracers of star formation (SF) in its 
disk, indicate that it is an actively star-forming galaxy.
It is amongst the first starburst galaxies observed with {\it 
XMM-Newton}, together with NGC\,253 (Pietsch et al. 2001), 
NGC\,3628 and M\,82 (unpublished). 
{\it Chandra} observations of NGC\,253, M\,82 and NGC\,3628 
were published by Strickland et al. (2000, 2002), Kaaret et al. 
(2001) and Matsumoto et al. (2001).

For an overview of papers related to investigations of galactic 
gaseous halos, refer to Dettmar (1992) and Dahlem (1997).
A description of the {\it XMM-Newton} observatory has been given
by Jansen et al. (2001).

\section{Observations and data reduction} 

Our observations of NGC\,1511 were carried out in 2000, July 7--8. 
The total on-source integration time of the European 
Photon Imaging Camera (EPIC) pn camera (Str\"uder et al. 2001)
is 40.0 ks. The pn camera performed the observation in extended
full frame mode and the two MOS cameras (Turner et al. 2001)
were put in full frame mode. For all cameras the thin filter
position was chosen, offering the highest possible transmission
for very soft X-ray emission.
The data were calibrated using the {\it XMM-Newton} Science Analysis 
System (SAS), version 5.3.3, and the Current Calibration File (CCF) 
version of 2002-10-10. Good time intervals were defined when the
total count rate, including background, was $< 1$ cts s$^{-1}$.
After removal of data from time intervals with high background 
count rates a net integration time of 32.1 ks remains (for the
pn camera). 

Corrections were applied to the data for the effects of gain 
variations of the preamplifier channels and charge transfer 
inefficiency (CTI), bad pixels, chip edges, split events etc.
and the counts were converted to the pulse-height invariant (PI) 
energy scale. Also, in order to fill the gaps 
in the EPIC pn camera images, images in the corresponding energy
bands were produced from the data of the two MOS cameras and all
images aligned and superimposed onto each other. The final images
presented here are a combination of data from all three EPIC 
cameras.
Photons with energies below 0.2 keV were not used in the data
analysis because of the calibration uncertainties in this very 
soft energy regime. Derived data products (images and spectra) 
were produced from the final event lists, after application of 
these corrections.

\section{X-ray Results and discussion}

In this paper we present only X-ray data from the European Photon 
Imaging Camera (EPIC; from the pn and both MOS cameras). 
No Reflection Grating Spectrometer (RGS, den Herder et al. 2001) 
results are used here, because the target is too faint and too 
extended for high-resolution X-ray spectroscopy.

\subsection{Soft X-ray imaging}

Our observations are the first dedicated X-ray observations of
NGC\,1511. Using the {\it ROSAT} All-Sky Survey (RASS) Browser we 
verified that it was not detected in the RASS.
In the data reduction we followed roughly the procedure adopted
by the {\it XMM-Newton} Survey Science Centre (SSC, Watson et al. 
2001) to split up the total passband into several energy bands. 
The only difference is that we chose to split the 0.5--2.0 keV band
into two bands, see Table~\ref{tab:banddef}. In one case we 
combined two SSC bands to show an image with energies ranging
from 2.0 keV to 7.5 keV.

\begin{table}[h!]
\begin{flushleft}
\leavevmode
\caption{XMM-Newton energy band definitions}
\label{tab:banddef}
\begin{tabular}{cccc}
\noalign{\hrule\smallskip}
  Number  & SSC energy range  &  Number  &   Energy ranges  \\
          &  definition (keV) &          &   used here (keV)   \\
\noalign{\hrule\smallskip}
   1      & 0.20--0.50 &   1      &  0.20--0.50     \\
   2      & 0.51--2.00 &   2a     &  0.51--1.30     \\
          &            &   2b     &  1.31--2.00     \\
   3      & 2.01--4.50 &   3      &  2.01--4.50     \\
   4      & 4.51--7.50 &   4      &  4.51--7.50     \\
   5      & 7.51--12.0 &   5      &  7.51--12.0     \\
\noalign{\smallskip\hrule}
\end{tabular}
\end{flushleft}
\end{table}

\begin{figure*}[t!]
\psfig{figure=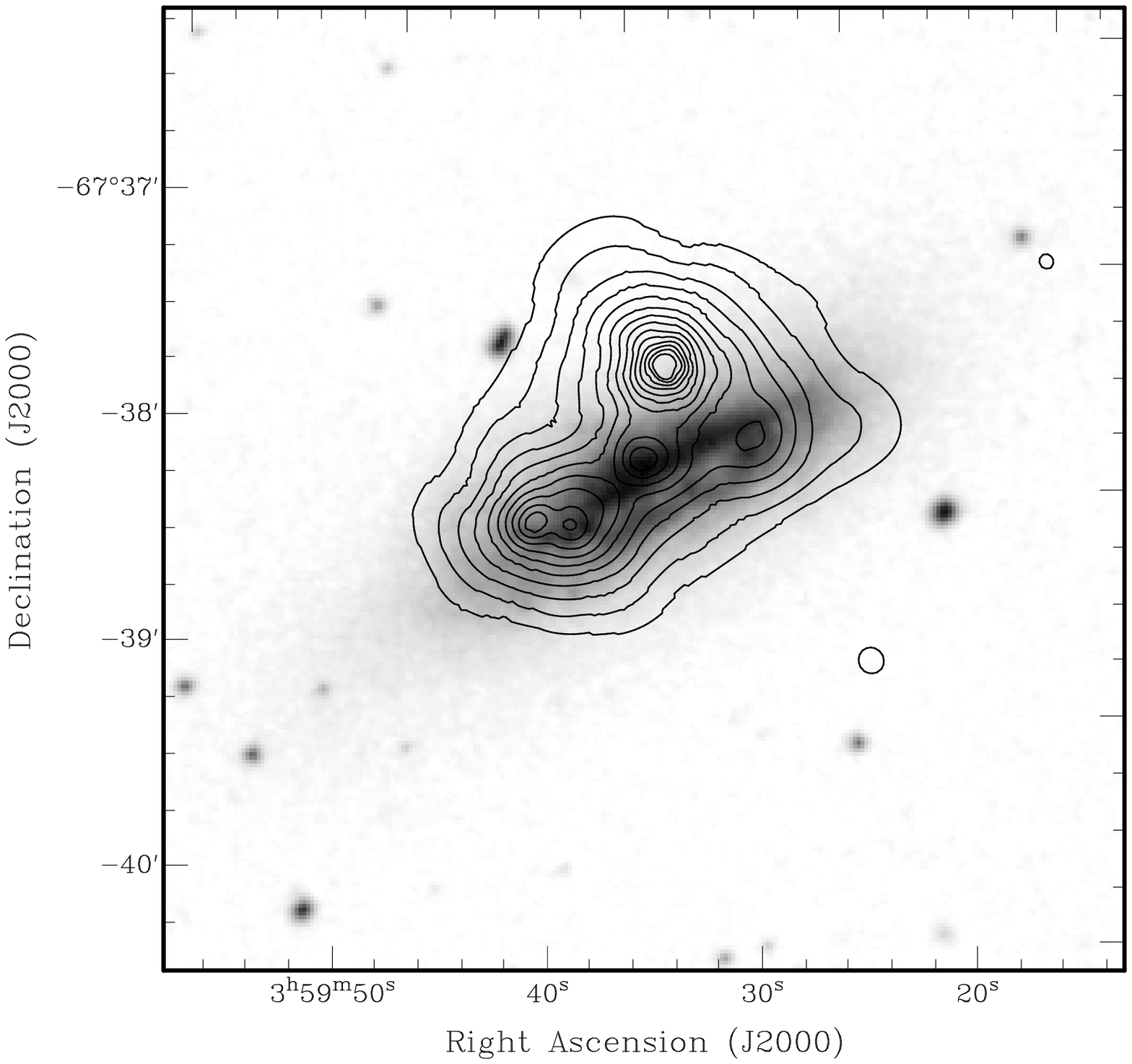,width=8.8cm}
\vspace*{-8.9cm}
\hspace*{9.0cm}
\psfig{figure=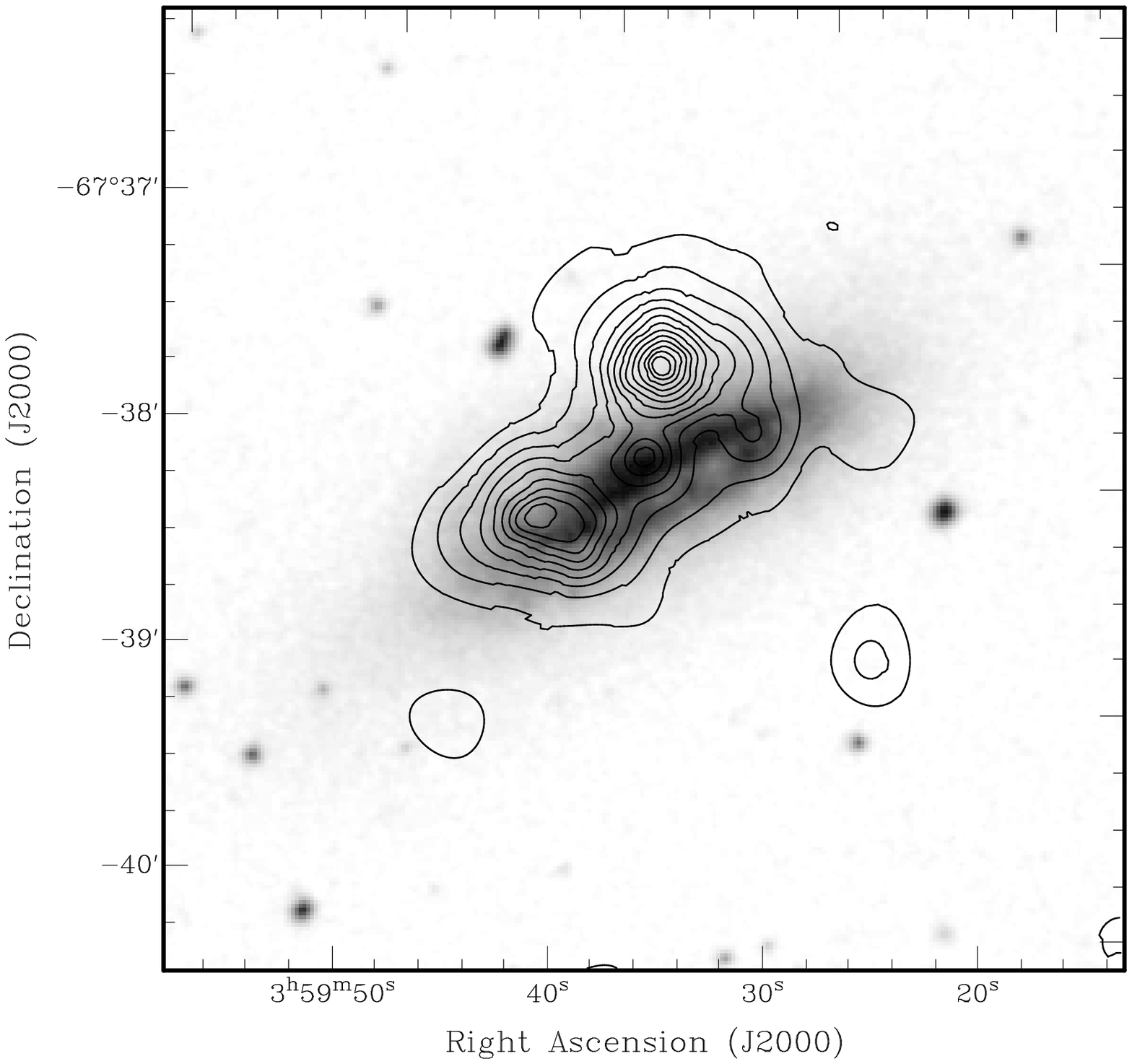,width=8.8cm}
\caption{
{\it XMM-Newton} 1.3--2.0 keV image (left), and 2.0--7.5 keV 
image (right) of NGC\,1511 on DSS-2. For details see text. The 
displayed field-of-view is $255''\times255''$, or $21.6\times 
21.6$ kpc, for an adopted distance towards NGC\,1511 of 17.5 
Mpc.
\label{fig:n1511im}
} 
\end{figure*}

In Fig.~\ref{fig:n1511im}\ we display X-ray images of NGC\,1511. 
The photon distributions which were received in the individual 
energy bands as defined above were smoothed with an adaptive 
filter so as to enhance low surface brightness features, while 
not smearing out the flux of strong point sources. Such images 
are not suitable for quantitative analyses.
The left panel is an overlay of the adaptively-smoothed {\it 
XMM-Newton} 1.3--2.0 keV image on a second generation Digitized 
Sky Survey (DSS-2) red image.
The right panel shows an overlay of the adaptively-smoothed 
{\it XMM-Newton} 2.0--7.5 keV image on the same DSS-2 image of 
NGC\,1511.
Contours start at the 2.5-$\sigma$ confidence level above the
background and are spaced by factors of $\sqrt{2}$. The 2.5-$\sigma$ 
levels of the images displayed here are: 
$1.14\,10^{-6}$ cts s$^{-1}$ arcmin$^{-2}$ (0.5--1.3 keV), 
$0.62\,10^{-6}$ cts s$^{-1}$ arcmin$^{-2}$ (1.3--2.0 keV) and 
$1.20\,10^{-6}$ cts s$^{-1}$ arcmin$^{-2}$ (2.0--7.5 keV).

The X-ray emission from NGC\,1511 comprises different components, 
namely hard point sources and also soft extended emission. 
Extended soft emission is visible across the disk of the galaxy,
with a maximum at the position of a prominent SF region on its
eastern edge. A weak secondary maximum is visible near the centre
of the galaxy. In addition, the soft X-ray emission distribution
is extended perpendicular to the disk plane.
The strong X-ray point source about $30''$ north of the centre 
of NGC\,1511, which we will call--according to the IAU 
convention--XMMU\,J035936.6--673736, does not have an optical 
counterpart in either our {\it XMM-Newton} Optical Monitor (OM, 
Mason et al. 2001) images or the DSS. If associated with 
NGC\,1511, it might be an ultra-luminous 
($L_{\rm X} = 1.18\,10^{40}$ erg s$^{-1}$) X-ray source (ULX, 
see e.g. Strickland et al. 2001, Foschini et al. 2002 and 
references therein). If not related to NGC\,1511, it could be
a background source, e.g. an AGN (see Sect.~\ref{par:ulx}).

The emission distribution will be discussed in more detail below,
in comparison with other wavebands.

\subsection{Integral wide-band X-ray spectroscopy}

\subsubsection{Spectral extraction}

Spectroscopy was performed based on the EPIC pn event list 
only. Extracting valid (filter \#{\it XMMEA\_EP}) single 
and double photon events from NGC\,1511, an integral source 
spectrum was obtained. A separate spectrum was extracted for 
XMMU\,J035936.6--673736. The source photon extraction region 
for NGC\,1511 is an ellipse with a major axis radius of $67''$, 
a minor axis radius of $50''$ and a position angle of $45^\circ$, 
centred on the nucleus of NGC\,1511. From this, a circular
region of radius $15''$ around $\alpha(2000)$=03:59:36.6, 
$\delta(2000)$=--67:37:36.8 was excised for separate spectral
analysis of the emission from XMMU\,J035936.6--673736. 
A second, elliptical region was defined near NGC\,1511, 
on a neighbouring pn chip, to determine the local 
background.\footnote{The background photon extraction region 
is an ellipse equal in size to that for the source photons, 
but centred at $\alpha(2000)$=03:59:05.40, 
$\delta(2000)$=--67:41:09.9.}       
Extracted spectra were grouped based both on a minimum number of 
counts (25) for each channel and on the maximum number of spectral 
channels sampling the pn energy resolution (3).
An analysis of the background-subtracted integral spectrum of 
NGC\,1511 over the energy range 0.2--12.0 keV was performed using 
Xspec v. 11.1.0.

\subsubsection{The need for a multi-component fit}

\begin{table*}[t!]
\caption{Spectral fit parameters for different models}
\label{tab:specfits}
\begin{flushleft}
\begin{tabular}{lccccccccc}
\noalign{\hrule\smallskip}
 ~(1) &  (2) & (3) & (4) & (5)  & (6) & (7) & (8) & (9) & (10) \\
&\multicolumn{3}{c}{Soft} & \multicolumn{3}{c}{Medium} 
  & \multicolumn{2}{c}{Hard} & \\
Model & $N_{\rm H}^a$(c) & $kT_1^b$ & $Z^c$ & $N_{\rm H}^a$(w) & $kT_2^b$
  & $Z^c$ & $N_{\rm H}^a$(h) & $\Gamma$ & $\chi^2_\nu$/d.o.f. \\
\noalign{\hrule\smallskip}
MP  & 4.3$^{+1.7}_{-0.1}$ & 0.24$^{+0.03}_{-0.05}$ & 1.0(f) & & &
    & 2.4$^{+2.5}_{-0.8}$ & 1.84$^{+0.25}_{-0.20}$ & 0.973/87 \\
MMP & 4.1$^{+2.2}_{-1.3}$ & 0.19$^{+0.04}_{-0.03}$ & 1.0(f) 
    & 1.8$^{+2.1}_{-1.8}$ & 0.59$^{+0.14}_{-0.10}$ & 1.0(f) 
    & 1.8$^{+2.1}_{-1.8}$ & 1.95$^{+0.22}_{-0.21}$ & 0.912/83 \\
\noalign{\smallskip\hrule}
\end{tabular}
\end{flushleft}
\noindent Notes to Table~\protect\ref{tab:specfits}:\\
M = Mekal plasma; P = power law; f = metallicity, $Z$, fixed at 1.0 $Z_\odot$.\\
$^a$ $N_{\rm H}$ in units $10^{21}$ cm$^{-2}$. \\
$^b$ $kT$ in units keV. \\
$^c$ $Z$ in units $Z_\odot$. \\
\end{table*}

Our first attempts, with single emission components plus foreground
absorption, for all different kinds of spectral models, provided 
statistically unacceptable fit results. 
Therefore, we subsequently tried a combination of a Mewe-Kaastra
plasma with Fe-L lines by Liedahl (``Mekal'') model (Mewe et al. 
1985, Mewe et al. 1986, Kaastra 1992) plus a power law, also 
allowing the two components to undergo different absorption. A 
thermal plasma was chosen to represent the diffuse emission 
component, while the power law approximates the expected spectrum 
of the point source(s), which, if associated with NGC\,1511, are 
most likely to be binaries, otherwise background AGN.
This improved the quality of the fit (see Table~\ref{tab:specfits}), 
but still some residual emission remained around 0.6, 1.0 and 2.0 
keV. 
Adding to the spectral model a second thermal component leads to a 
statistically marginally better fit, but improved visibly the fit 
in the areas that showed residuals before. Thus, the spectrum of 
NGC\,1511 clearly requires a multi-component fit. However, the one 
presented here is not necessarily the only possible good fit to the 
spectrum (Table~\ref{tab:specfits}).
At this time, we can only state that a single thermal component 
plus power law spectral model cannot explain the entire soft X-ray 
spectrum of NGC\,1511, demonstrating the complexity of the emission 
spectrum. The nature of the additional third component is as yet
uncertain.

\subsubsection{Best-fitting spectral model for NGC\,1511}

\begin{figure}[h!]
\psfig{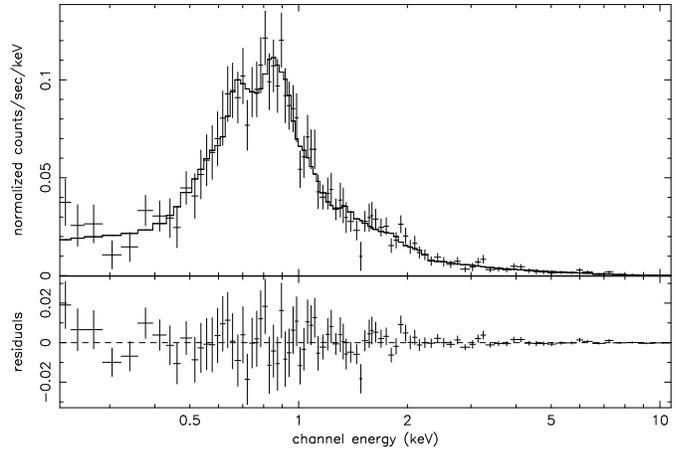}
\caption{
Total {\it XMM-Newton} EPIC pn spectrum of NGC\,1511. The spectral 
model fitted to the data includes two thermal plasma components 
plus a power law, all with absorption by foreground gas (model
``MMP'' in Table~\protect\ref{tab:specfits}). In the lower panel 
the residuals after subtraction of the spectral model are 
presented.
\label{fig:n1511spec}
} 
\end{figure}

The three-component fit with two thermal components described in 
Table~\ref{tab:specfits}\ is displayed in Fig.~\ref{fig:n1511spec}.
Two thermal components, with their metallicities fixed at 1.0 
$Z_\odot$, were fitted. Fitting the metallicities of the thermal 
components as a free fit parameter did not statistically improve
the fit quality and the $Z$ values are not constrained sensibly.
Therefore, this fit is not shown here.
In the fit with fixed metallicities, which we consider the
simplest best-fitting model, the relative contribution of the 
two thermal components and the power law to the total flux are 
12\% (0.19 keV), 11\% (0.59 keV) and 77\%, respectively. 
%
%
These relative flux contributions conceal the fact that diffuse 
thermal emission dominates the total emission in the soft regime, 
while at hard energies the power-law component is dominant.
We have verified by means of spatially resolved spectroscopy
of subregions that the thermal emission corresponds with the 
extended soft component visible in our images.
The best-fitting absorbing column densities are higher than the 
Galactic foreground ($N_{\rm H} = 0.5\,10^{21}$ cm$^{-2}$, Dickey \& 
Lockman 1990), suggesting additional internal absorption in 
NGC\,1511.
Based on this fit, we determine a total 0.2--12 keV flux of 
NGC\,1511 of $f_{\rm X} = 3.04\,10^{-13}$ erg s$^{-1}$ cm$^{-2}$.
This X-ray flux translates to a 0.2--12 keV X-ray luminosity 
of NGC\,1511 of $L_{\rm X} = 1.11\,10^{40}$ erg s$^{-1}$. 
>From this value a far-infrared--to--X-ray luminosity ratio of 
log$(L_{FIR}/L_{\rm X}) = 3.62$ follows, which is typical for 
starburst galaxies (Heckman et al. 1990; Read \& Ponman 2001). 
The total {\it diffuse} soft X-ray luminosity of NGC\,1511 is
$L_{\rm X} = 2.55\,10^{39}$ erg s$^{-1}$. This leads to a ratio 
of log$(L_{\rm FIR}/L_{\rm X,diff}) = 4.25$. The amount of 
diffuse soft X-ray emission in NGC\,1511, compared to other 
galaxies, is thus relatively low (Read \& Ponman 2001; their 
Fig.~5, central panel).
The temperatures of the thermal plasma components are in
line with earlier measurements of other starburst galaxies, 
based on {\it ROSAT} and {\it ASCA} data (e.g. Dahlem et al. 
1998; Weaver et al. 2000 and references therein). Components 
at about 0.2--0.3 keV and 0.6 keV were found by us in several 
starburst galaxies, with the 0.2--0.3 keV component dominant 
in the halos, while the hotter component is normally observed 
from gas in the disks. Similar results were also obtained 
based on previously published data from {\it Chandra} and 
{\it XMM-Newton} (see references in Sect.~1).

\subsubsection{Model fit to the spectrum of XMMU\,J035936.6--673736}
\label{par:ulx}

The best-fitting model spectrum of XMMU\,J035936.6--673736 (not 
displayed) is a power-law with $\Gamma=1.92^{+0.14}_{-0.11}$ 
and an absorbing column density of $N_{\rm H} = 2.5^{+1.0}_{-0.3}\,10^{21}$ 
cm$^{-2}$. The 0.2--12 keV X-ray flux of this source is
$f_{\rm X} = 3.22\,10^{-13}$ erg s$^{-1}$ cm$^{-2}$.
The fitted moderately low absorbing column density makes it
unlikely that this source is a background AGN, which one
would expect to have much higher intrinsic absorption.

\begin{figure*}[t!]
\psfig{figure=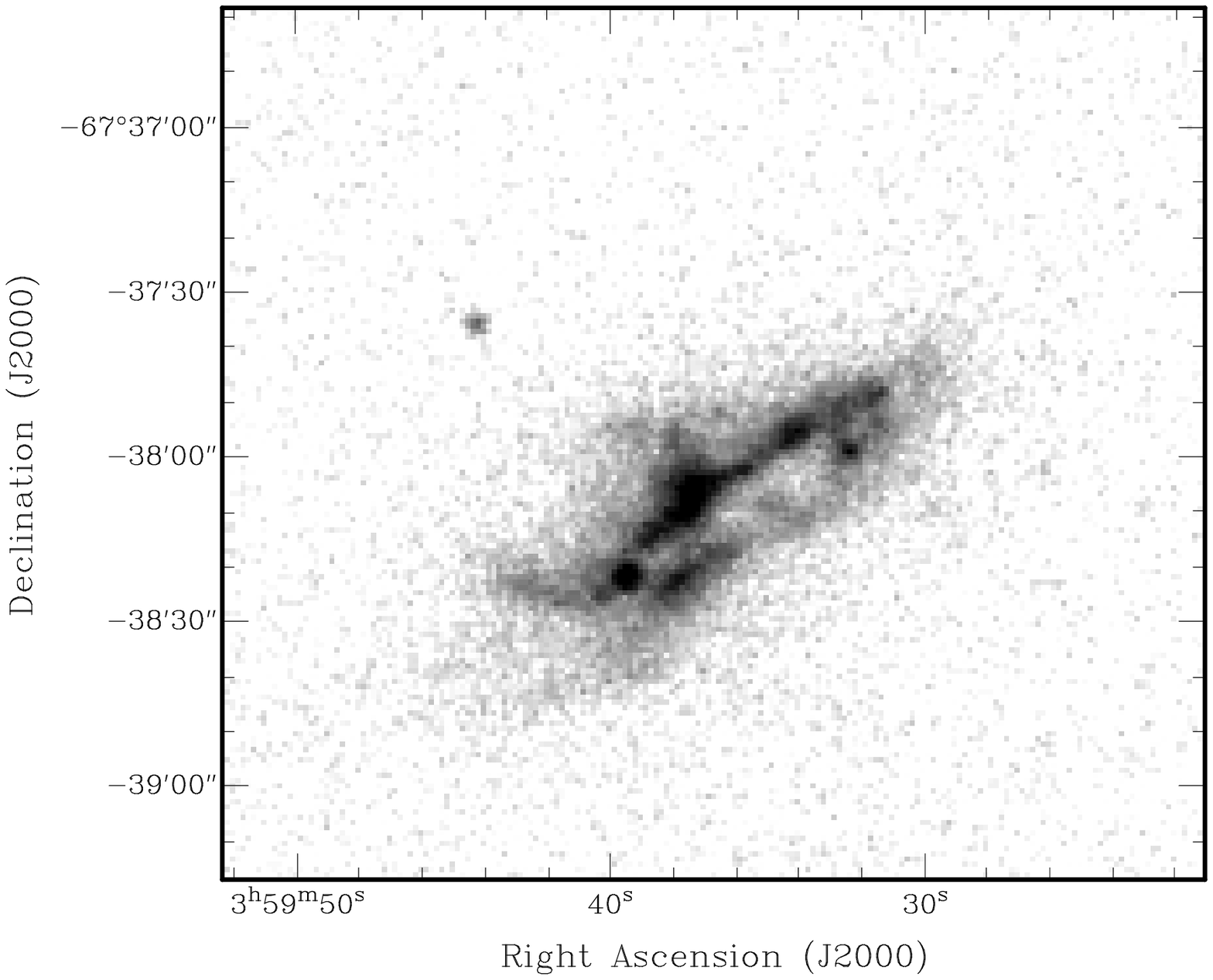,width=8.8cm}
\vspace*{-8.9cm}
\hspace*{9.0cm}
\psfig{figure=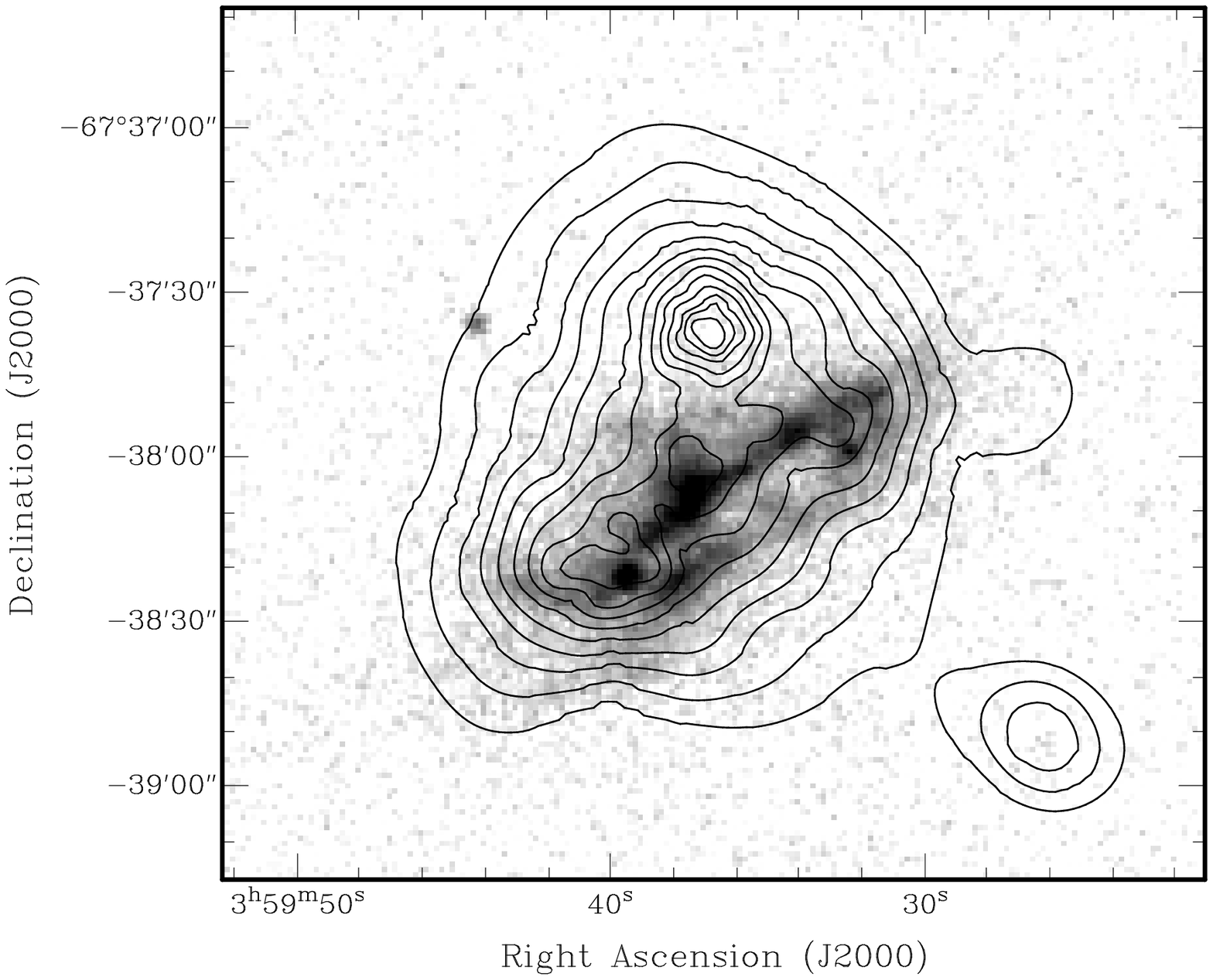,width=8.8cm}
\caption{
{\it XMM-Newton} OM image obtained with the UVM2 filter, with 
a central wavelength of about 230 nm (left) and the same image, 
with the soft X-ray band 0.5--1.3 keV overlaid (right). The 
displayed field-of-view is $180''\times160''$, or 
$15.4\times 13.6$ kpc.
\label{fig:n1511uvm2}
} 
\end{figure*}

\subsection{UV imaging}

The OM onboard {\it XMM-Newton}, with its three UV filters in 
addition to a range of optical filters, opens a second window 
on the targets observed that is not accessible from the ground, 
in the UV. As an example, we present in the left panel of 
Fig.~\ref{fig:n1511uvm2} a 230 nm image of NGC\,1511 obtained 
with the UVM2 filter.

This image exhibits the UV emission from massive stars, which 
are the primary sources of energy heating the warm ionized gas 
in NGC\,1511. This is the reason for the global correspondence 
between the emission maxima in the UV and H$\alpha$ images. 
However, it is clear from a comparison of the two frames
(Fig.~\ref{fig:n1511uvm2}\ [left panel] and Fig.~\ref{fig:n1511hair}\ 
[right panel]) that there is more diffuse H$\alpha$ outside 
the \hii -regions. 
The presence of such massive stars is yet another indicator 
of NGC\,1511's activity, as already suggested by its 
$L_{\rm X}/L_{\rm FIR}$ ratio and also by the presence of 
diffuse halo emission in various wavebands.

In the overlay with the EPIC 0.5--1.3 keV soft X-ray image in 
the right panel of Fig.~\ref{fig:n1511uvm2}\ one can see that 
there is basically no spatial correspondence between UV and 
X-ray emitters in NGC\,1511.

The UV point source south-east of the centre of NGC\,1511 is 
probably an unrelated foreground star. It also shows up on an 
R-band image by Lehnert \& Heckman (1995). 
%
Its position does not coincide with that of SN1935C or Nova Hydri
1935 (van den Bergh \& Hazen 1988). 

An OM image obtained with the UVW2 filter (central wavelength 
about 210 nm), exhibits the same general emission distribution as 
Fig.~\ref{fig:n1511uvm2}, however with a lower signal-to-noise 
ratio. Therefore, it is not displayed here.

\section{A multi-wavelength picture of NGC\,1511}

The observations described here are part of a large project
aiming at understanding the nature of gaseous halos around
actively star-formating galaxies.

\begin{figure*}[t!]
\psfig{figure=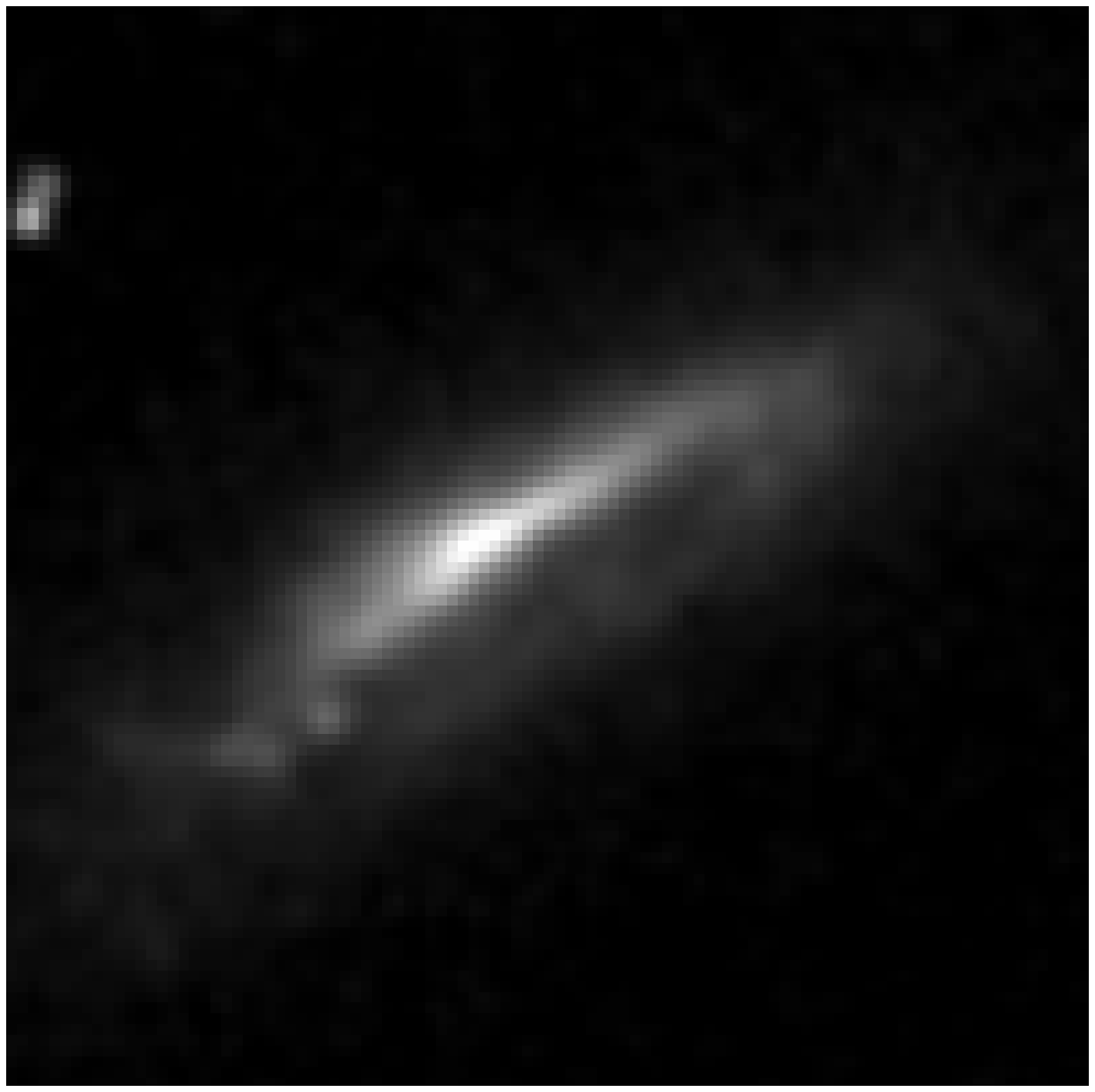,width=8.8cm,angle=0}
\vspace*{-7.5cm}
\hspace*{9.0cm}
\psfig{figure=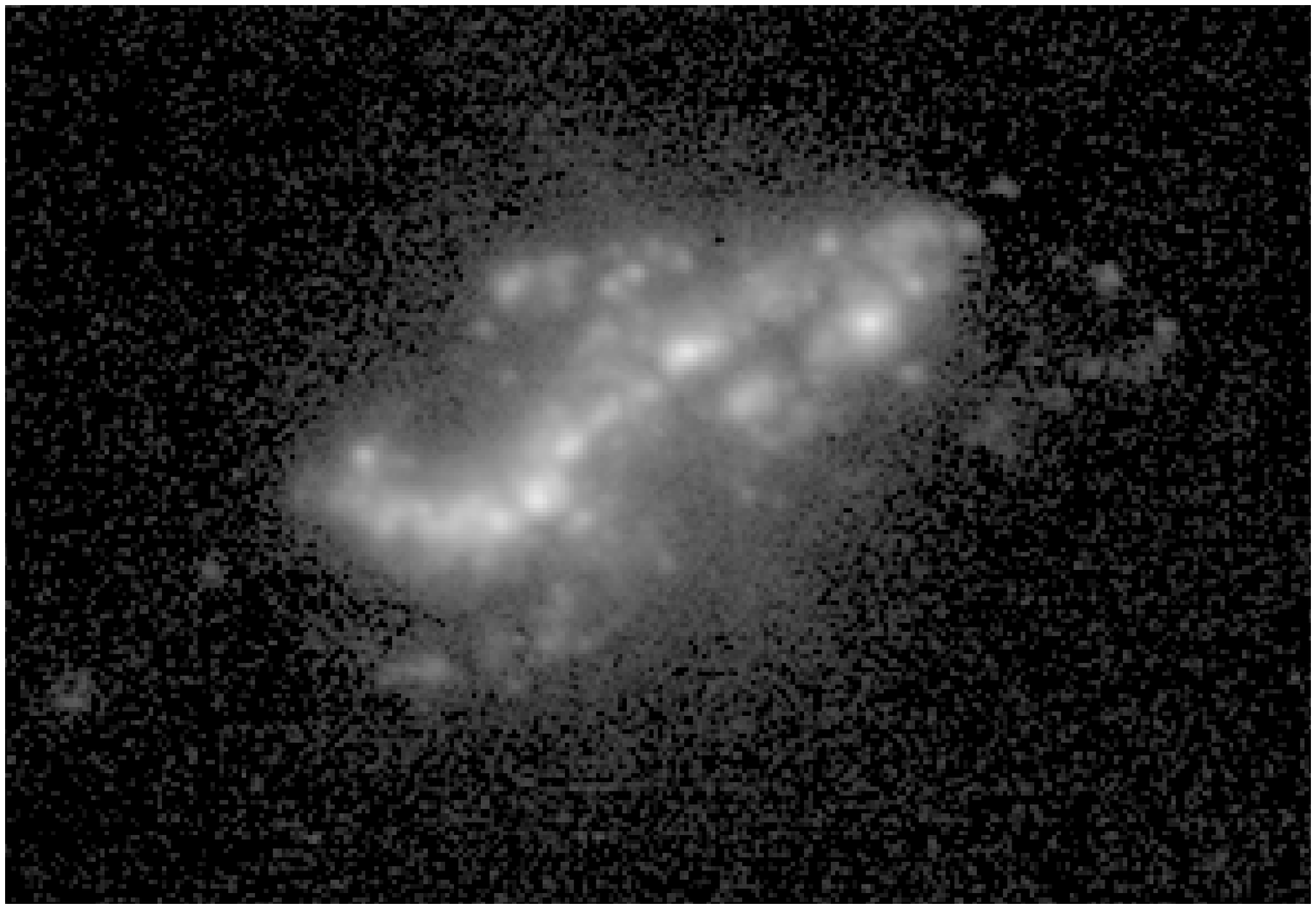,width=8.8cm,angle=270}
\vspace*{2cm}
\caption{
{\bf Left panel:}
2MASS J-H-K$_{\rm s}$ image of NGC\,1511. North is to the 
top, east to the left. The displayed field of view is 
$101''\times101''$, or $8.6\times8.6$ kpc.
{\bf Right panel:}
H$\alpha$ image of NGC\,1511 published by Lehnert \& Heckman 
(1995). In their contour representation of this image the 
disturbed distribution of H$\alpha$ emission could not be 
seen as clearly as in this logarithmically scaled display. 
North is to the top, east to the left. The displayed field 
of view is $161''\times110''$, or $13.7\times9.3$ kpc.
\label{fig:n1511hair}
} 
\end{figure*}

In the case of NGC\,1511 it is interesting to compare the {\it 
XMM-Newton} images with others obtained in different wavebands. 
Especially a comparison of the OM image with H$\alpha$ and 2
Micron All-Sky Survey (2MASS; Skrutskie 2001) near-infrared 
frames is useful. 
Both, the H$\alpha$ frame by Lehnert \& Heckman (1995) and the
2MASS J-H-K$_s$-band composite image are displayed in 
Fig.~\ref{fig:n1511hair}.

\subsection{Extraplanar emission}

The X-ray disk emission follows roughly that of H$\alpha$ and 
radio continuum (Lehnert \& Heckman 1995; Dahlem et al. 2001).
In particular the correspondence between soft X-ray and radio
emission is obvious. The axial ratio of both emission distributions
is much smaller than that of the optically visible galaxy and there
are two emission maxima in the disk, one near the centre and one
in the south-eastern part of the disk.
NGC\,1511's disk is disturbed and the disk plane not well-defined, 
as visible e.g. in the H$\alpha$ by Lehnert \& Heckman (1995). 
Also the NIR 2MASS images suggest disturbances of the stellar 
distribution. This makes it difficult to assess whether emission 
arises from inside or beyond the disk. 
Still, the soft X-ray emission to the north of the central region 
comes from an area beyond the optically visible galaxy, thus 
possibly from the halo. This corroborates our earlier finding 
(e.g. Dahlem 1997) that, once gaseous halos are detected around 
actively star-forming spirals, different phases of the ISM are 
present. H$\alpha$, radio continuum and diffuse soft X-ray 
emission (i.e., warm ionized gas, cosmic rays, magnetic fields
and hot ionized gas) are found to be associated with each other, 
although not necessarily directly correlated.

\subsection{Disturbances of NGC\,1511}

The H$\alpha$ frame shows quite impressively how strong the
starburst activity in NGC\,1511 is and also how disturbed 
the galaxy is. This disturbance is most likely caused by its
companions, NGC\,1511a and NGC\,1511b, which are observed
at projected distances from NGC\,1511 of 56.7 kpc and 38.7 
kpc, respectively, with very small relative recession velocities 
of only 38(25) km s$^{-1}$ with respect to NGC\,1511.
Using a high-contrast display of the corresponding red DSS-2 plate 
one can see a tidal tail connected to NGC\,1511 (not visible in
Fig.~\ref{fig:n1511im}, because with such a grey-scale no contours 
could be seen), from the south-eastern end of the galaxy pointing
roughly northwards, suggesting an ongoing tidal interaction.

The starburst nature of NGC\,1511 is also corroborated by an 
updated radio spectral index calculation that we performed, 
using our earlier measurements (Dahlem et al. 2001), plus a 
4.85 GHz measurement by Gregory et al. (1994) and an 843 MHz 
flux density measurement from the Sydney University Molonglo 
Sky Survey (SUMSS; Bock et al. 1999). 
$S_{4.85GHz} = 76$ mJy and $S_{0.843GHz} = 237$ mJy lead to a 
new, improved value of $\bar\alpha\ = -0.66\pm0.08$ 
($S\propto\nu^{\bar\alpha}$), a value typical of actively 
star-forming galaxies (Condon 1992). This new value is, within the 
ample error bar of our earlier calculations, consistent with 
these. The Molonglo Telescope 843 MHz map (not displayed) also 
corroborates the emission distribution of our 1.43 GHz and 2.45 
GHz maps (Dahlem et al. 2001) with a low major-to-minor axial 
ratio, indicating the presence of a synchrotron radio halo.

The 2MASS image suggests that NGC\,1511's appearance as a
high-inclination ring system, as seen in the UV image in
Fig.~\ref{fig:n1511uvm2}, might be caused not by such an
emission distribution, but by absorption by a massive dust
lane that is optically thick even in the NIR, creating a
central emission depression.

This multi-wavelength approach demonstrates that NGC\,1511 is 
a heavily disturbed starburst galaxy, which implies that it is 
not a good candidate for investigations of the dependence of 
its gaseous halos' properties on the level and distribution of 
star formation in the underlying disk, for which one would 
prefer to study relatively undisturbed systems. 
Processes other than energy input from stellar sources (stellar 
winds and supernovae; see Leitherer \& Heckman 1995) might
influence the gas motions vertical to the disk, such as e.g.
a gravitational potential modified by a tidal interaction with
a companion galaxy. 
Still, NGC\,1511 is one of a small number of starburst galaxies 
in which various components of the ISM are observed, within the 
same volume, above their disk planes. A significant contribution 
to the heating of its ISM from a hidden AGN can be excluded based 
on our X-ray data due to the absence of a hard central source, 
from optical spectra because of the absence of typical AGN emission 
lines and from our radio data, which also do not show any signatures
of AGN activity at the centre.

\subsection{Related observations}

NGC\,1511, amongst other objects, is also being observed by 
us in \hi\ line emission with the Australia Telescope Compact 
Array to study the properties of its neutral atomic gas. This 
will lead to new information about the intrinsic absorbing gas 
distribution and its kinematics. It will, for example, also
help to elucidate the issue whether the observed disturbances
of NGC\,1511's disk can be explained by interactions with its
companions, NGC\,1511a and NGC\,1511b.

{\it XMM-Newton} observations of several more nearby starburst 
galaxies, including NGC\,1808, NGC\,3628 and NGC\,4666, have 
been conducted by us. These, together with data from the {\it 
XMM-Newton} archive, will be used to study systematically and 
in a self-consistent manner the soft X-ray emission properties 
of galaxies with kpc- to galaxy-scale outflows from their disks.

\vspace{0.5cm}

\noindent{\it Acknowledgements:} 
We thank Dr. M. Lehnert for making available to us his H$\alpha$
and R-band images of NGC\,1511 in digital form. 
%
%
This work is based on observations obtained with {\it XMM-Newton}, 
an ESA science mission with instruments and contributions directly 
funded by ESA Member States and the USA (NASA).
This research has made use of the NASA/IPAC Infrared Science 
Archive, which is operated by the Jet Propulsion Laboratory, 
California Institute of Technology, under contract with the 
National Aeronautics and Space Administration.
The Digitized Sky Surveys were produced at the Space Telescope
Science Institute under U.S. Government grant NAG W-2166. The
images of these surveys are based on photographic data obtained 
using the Oschin Schmidt Telescope on Palomar Mountain and the 
UK Schmidt Telescope. The plates were processed into the present
compressed digital form with the permission of these institutions. 


\end{document}